\documentclass{PoS}
\usepackage{amsmath}

\newcommand{\ket}[1]{\vert #1\rangle}
\newcommand{\bra}[1]{\langle #1\vert}
\newcommand{\braket}[1]{\langle #1\rangle}

\newcommand{\p}[1]{\left( #1 \right)} 
\newcommand{\s}[1]{\left[ #1 \right]} 

\title{
  \begin{picture}(0,0)(0,0)%
    \put(300,75){\makebox(0,0)[l]{\textnormal
        {\normalsize KEK-CP-340}
      }}
  \end{picture}Decay constants and spectroscopy of mesons in lattice QCD using domain-wall fermions
}

\ShortTitle{Decay constants of mesons using domain-wall fermions}

\author{JLQCD Collaboration: \speaker{B. Fahyz}$^a$\footnote{Email:bfahy@post.kek.jp},
  G. Cossu$^{a}$, S. Hashimoto$^{a,b}$, T. Kaneko$^{a,b}$, J. Noaki$^{a}$, M. Tomii$^{b}$\\
  \llap{$^a$} High Energy Accelerator Research Organization (KEK), Ibaraki 305-0801, Japan
  Address\\
  \llap{$^b$} Department of Particle and Nuclear Science, SOKENDAI (The Graduate University for
  Advanced Studies), Ibaraki 305-0801, Japan
}

\abstract{We report results of masses and decay constants of
  light and charmed pseudo-scalar mesons using lattice QCD with
  M\"obius domain-wall fermions. Using this formulation we are
  able to compute pseudo-scalar decay constants through the
  pseudo-scalar density operator as well as with the
  axial-vector current.  Results are shown from several lattice
  spacings and pion masses between 230 MeV and 500 MeV. We
  present an analysis of these results at different quark masses
  to show the chiral properties of the light mesons masses and
  decay constants.  }

\FullConference{The 33rd International Symposium on Lattice Field Theory\\
  14 -18 July 2015\\
  Kobe International Conference Center, Kobe, Japan*}

\begin{document}

\section{Introduction}

The JLQCD collaboration has recently produced ensembles of lattice
configurations with fine lattice spacings and good chiral
symmetry. Lattice simulations of 2+1-flavor QCD were performed using
the M\"obius domain-wall fermions~\cite{mobius} with tree-level
Symanzik gauge action. Table \ref{tab:lattices} shows the $15$ gauge
ensembles generated~\cite{finelattice}. These lattices have lattice
spacings of $1/a \approx 2.4$, $3.6$, and  $4.5\text{ GeV}$ with
pion masses between $230$ MeV and $500$ MeV. For each ensemble,
$10,000$ molecular dynamics (MD) times were run after
thermalization.

Using domain-wall fermions the Ginsparg-Wilson relation is only
approximate. The violation of the Ginsparg-Wilson relation is given by
the residual mass. The M\"obius representation as well as using stout
link-smearing~\cite{stout} make the residual mass small of
$\mathcal{O}(1\text{ MeV})$ on the coarsest ($\beta=4.17$) lattices and
$<0.2\text{ MeV}$ on the finer lattices~\cite{residual}. Good chiral
symmetry enables simpler renormalization such as $Z_V = Z_A$, and
simplifies the calculation of the pseudo-scalar decay constants directly
utilizing the PCAC relation.

With chiral symmetry preserved, observables can be used to compare
lattice QCD results to those of Chiral Perturbation Theory (ChPT). In
this work we present results of masses and decay constants of light
and charmed pseudo-scalar mesons. These measurements are used to
determine the low energy constants (LEC) in ChPT. Additionally the
fine lattice spacing enables exploration of charm physics with
manageable cutoff effects.

In this work the scale for these lattices was determined from Wilson
flow using $t_0$. We use the value from \cite{scale} as an input. The determination was done using
a linear extrapolation to the physical point in $m_\pi^2$ as well
as an interpolation of the strange quark mass to reproduce the physical
$(M^\text{phys}_{s\bar{s}})^2 = 2M_K^2 - M_\pi^2$. The fitted parameters to describe
the mass dependence for the two smaller $\beta$ values were then used
to determine the scale on our finest lattice. The results for $a^{-1}$
on each $\beta$ are listed in Table~\ref{tab:lattices}.

\begin{table}[h]
  \centering
  \begin{tabular}{|l|l|r|r|r|r|r|}
    \hline
    Lattice Spacing                    & $L^3\times T$                                 & $L_5$ & $a m_{ud}$ & $a m_s$ & $  m_\pi \text{ [MeV]} $ & $ m_{\pi}L $ \\
    \hline
    $\beta = 4.17,$                      & $ 32^3\times64$ $(L=2.6 \text{ fm})$ & 12    & 0.0035   & 0.040   & 230                    & 3.0          \\
    $a^{-1}=2.453(4)\text{ GeV}$    &                                                 &       & 0.0070   & 0.030   & 310                    & 4.0          \\
                                       &                                                 &       & 0.0070   & 0.040   & 310                    & 4.0          \\
                                       &                                                 &       & 0.0120   & 0.030   & 400                    & 5.2          \\
                                       &                                                 &       & 0.0120   & 0.040   & 400                    & 5.2          \\
                                       &                                                 &       & 0.0190   & 0.030   & 500                    & 6.5          \\
                                       &                                                 &       & 0.0190   & 0.040   & 500                    & 6.5          \\ \cline{2-7}
                                       & $48^3\times96 $ $ (L=3.9 \text{ fm})$             &  12     & 0.0035   & 0.040   & 230                    & 4.4          \\
    \hline
    $\beta= 4.35,$               & $48^3\times 96 $ $(L=2.6 \text{ fm})$           & 8     & 0.0042   & 0.018   & 300                    & 3.9          \\
    $a^{-1}=3.610(9)\text{ GeV}   $ &                                                 &       & 0.0042   & 0.025   & 300                    & 3.9          \\
                                       &                                                 &       & 0.0080   & 0.018   & 410                    & 5.4          \\
                                       &                                                 &       & 0.0080   & 0.025   & 410                    & 5.4          \\
                                       &                                                 &       & 0.0120   & 0.018   & 500                    & 6.6          \\
                                       &                                                 &       & 0.0120   & 0.025   & 500                    & 6.6          \\
    \hline
    $\beta = 4.47,$              & $64^3\times128 $ $(L=2.7  \text{ fm}) $          & 8     & 0.0030   & 0.015   & 280                    & 4.0          \\
    $a^{-1} = 4.496(9) \text{ GeV}$ &                                                 &       &          &         &                        &              \\
    \hline
  \end{tabular}
  \caption{Parameters of the JLQCD gauge ensembles used in this work. Pion masses are rounded to the nearest $10$ MeV. The ensemble with $m_\pi L \approx 3.0$ is excluded in all analysis below to avoid possible finite volume effects.  \label{tab:lattices}}
\end{table}

\section{Computation of observables}

Pseudo-scalar correlators were produced utilizing our QCD software
package Iroiro++~\cite{iroiro}. These correlators were computed on
$200$ gauge configurations separated by $50$ MD times and from two
source locations, producing $400$ measurements of the light
correlators and $300$ measurements of heavy correlators for each
ensemble, except for the $\beta=4.17$ ensemble on the larger volume,
which has $600$ light and $400$ heavy measurements. Correlators were
produced with unsmeared point sources as well as smeared sources using
Gaussian smearing, and the same point and smeared operators are used
also for the sinks. Gaussian smearing is defined by the operator
$(1-(\alpha/N) \Delta)^N$ where $\Delta$ as the Laplacian and in this
work the parameters $\alpha= 20.0$ and $N=200$ were used.

The amplitudes of the unsmeared local operators are required to
compute the decay constants. Two-point correlation functions of the
form $\braket{P^L(x){P^G}^\dagger(0)}$ were fit simultaneously with correlators
$\braket{P^G(x){P^G}^\dagger(0)}$ where $L$ indicates an unsmeared local
operator while $G$ denotes Gaussian smeared operators.
The two-point correlation functions were fit to the functional form
\begin{align}
  \label{eq:amp}
  C = \underbrace{\frac{1}{2m_\pi} \bra{0}P \ket{\pi}\bra{\pi}P^\dagger\ket{0}}_{A_{PP}} \p{e^{-m_\pi t}+e^{-m_\pi (N_t-t)}}
\end{align}
for large $t$ to determine the masses and amplitudes where $P$ is
either $P^L$ or $P^G$. The matrix element of $\bra{0}P\ket{\pi}$ of
the unsmeared operator $P^L$ can be reconstructed from the
simultaneous fit of $\braket{P^L(x)P^G(0)}$ and
$\braket{P^G(x)P^G(0)}$.  The decay constants are calculated by
utilizing the axial Ward-Takahashi identity
${Z_A\partial_\mu A_\mu=(m_{q_1}+m_{q_2})P}$, where $A_\mu$ is the
lattice axial current, and $m_q$'s are the quark masses of the
pseudo-scalar meson of interest. This leads to the formula for $f_P$,
\begin{align}
  \label{eq:fpi}
  f_P = (m_{q_1} + m_{q_2})\sqrt{\frac{2A_{PP}}{m_\pi^3}},
\end{align}
which does not rely on the  renormalization constant $Z_A$. The mass
$m_q$ used is the bare quark mass plus the residual mass. We use the convention $F_\pi = f_\pi/\sqrt{2}$.


\section{Pion masses and decay constants}

Our measurements of the pion masses and decay constants for ensembles
at different bare light quark masses allow us to investigate the
consistency with $SU(2)$ ChPT. The quark mass
dependence of $M_\pi$ and $F_\pi$ at next-to-next-to-leading
order~\cite{PhysRevD.90.114504} is
\begin{align}
  \frac{M_\pi^2}{\bar{m}_q } &= 2B\s{1-\frac{1}{2}x \ln\frac{\Lambda_3^2}{M^2} + \frac{17}{8} x^2\p{\ln\frac{\Lambda_M^2}{M^2}}^2 + k_M x^2 +\mathcal{O}\p{x^3} }, \label{mx-expand}\\
  F_\pi &= F\s{1 + x \ln\frac{\Lambda_4^2}{M^2}-\frac{5}{4} x^2\p{\ln\frac{\Lambda_F^2}{M^2} }^2 + k_F x^2 +\mathcal{O}\p{x^3}}.
          \label{fx-expand}
\end{align}
These are expanded using the parameter $x = M^2/(4\pi F)^2$ where
$M^2 = B (\bar{m}_q +\bar{m}_q ) = 2\bar{m}_q \Sigma/F^2$. $\bar{m}_q$
is the appropriately renormalized quark mass, where the renormalization
factor is discussed in \cite{renormalization}. The parameters
$\Lambda_3$ and $\Lambda_4$ are related to the effective coupling
constants of ChPT through $\bar{\ell}_n = \ln
{\Lambda_n^2/M_\pi^2}$. $\Lambda_M$ and $\Lambda_F$ are linear
combinations of different $\Lambda_n$'s~\cite{PhysRevD.90.114504}.

The chiral expansions above are fit to the data for $F_\pi$ and
$M_\pi^2/\bar{m}_q $ simultaneously at both NLO and NNLO. At NLO only
terms up to $\mathcal{O}{\p{x^2}}$ in (\ref{mx-expand}) and
(\ref{fx-expand}) are included leaving the free parameters
$F, B,\Lambda_3$, and $\Lambda_4$. For NNLO there are the additional
free parameters $k_M$, and $k_F$, while the values of $\Lambda_{1}$
and $\Lambda_{2}$ were fixed to the phenomenological value from
\cite{colangelo}.

\begin{figure}[t]
  \centering
  \includegraphics[width=0.49\textwidth]{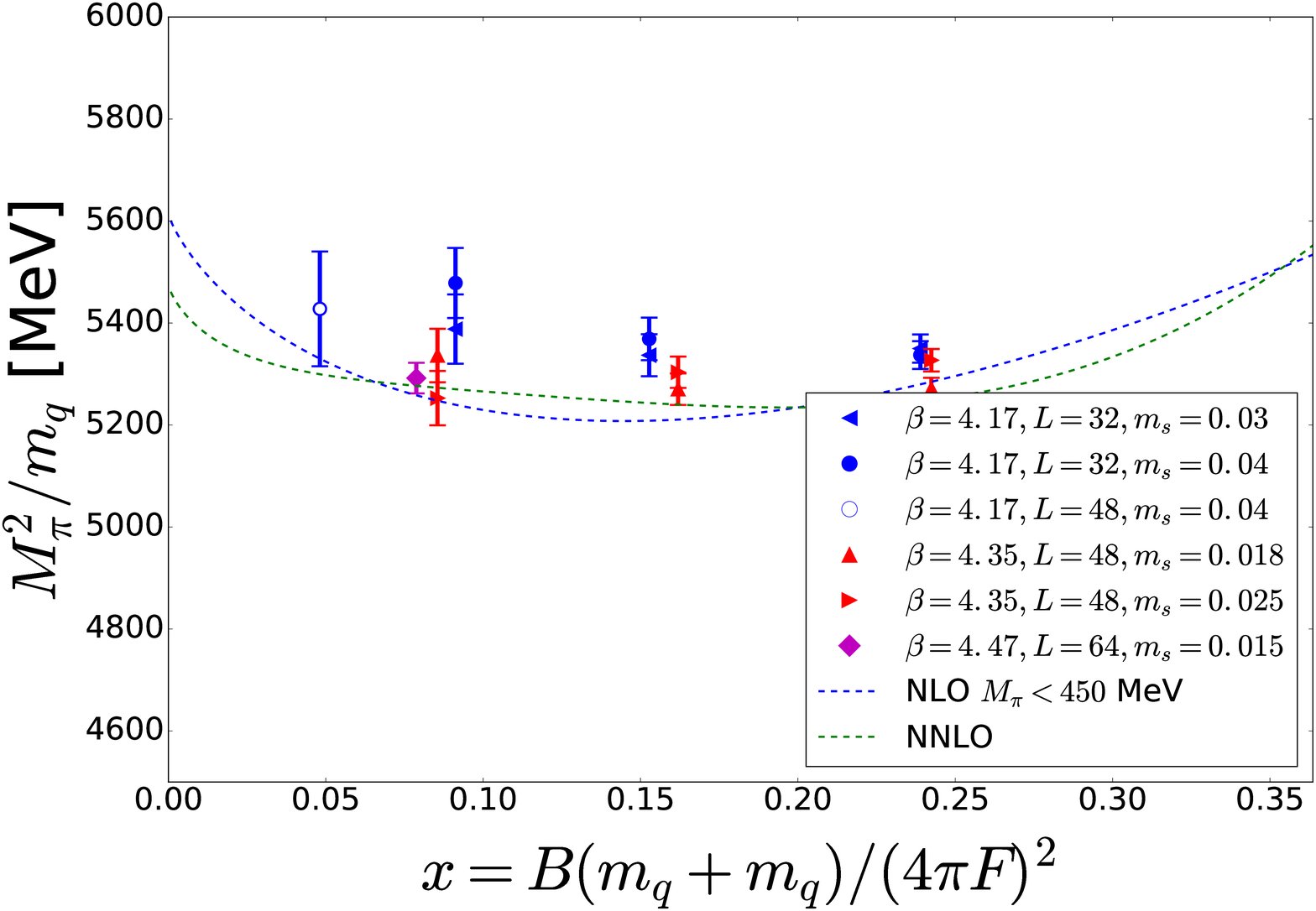}
  \includegraphics[width=0.49\textwidth]{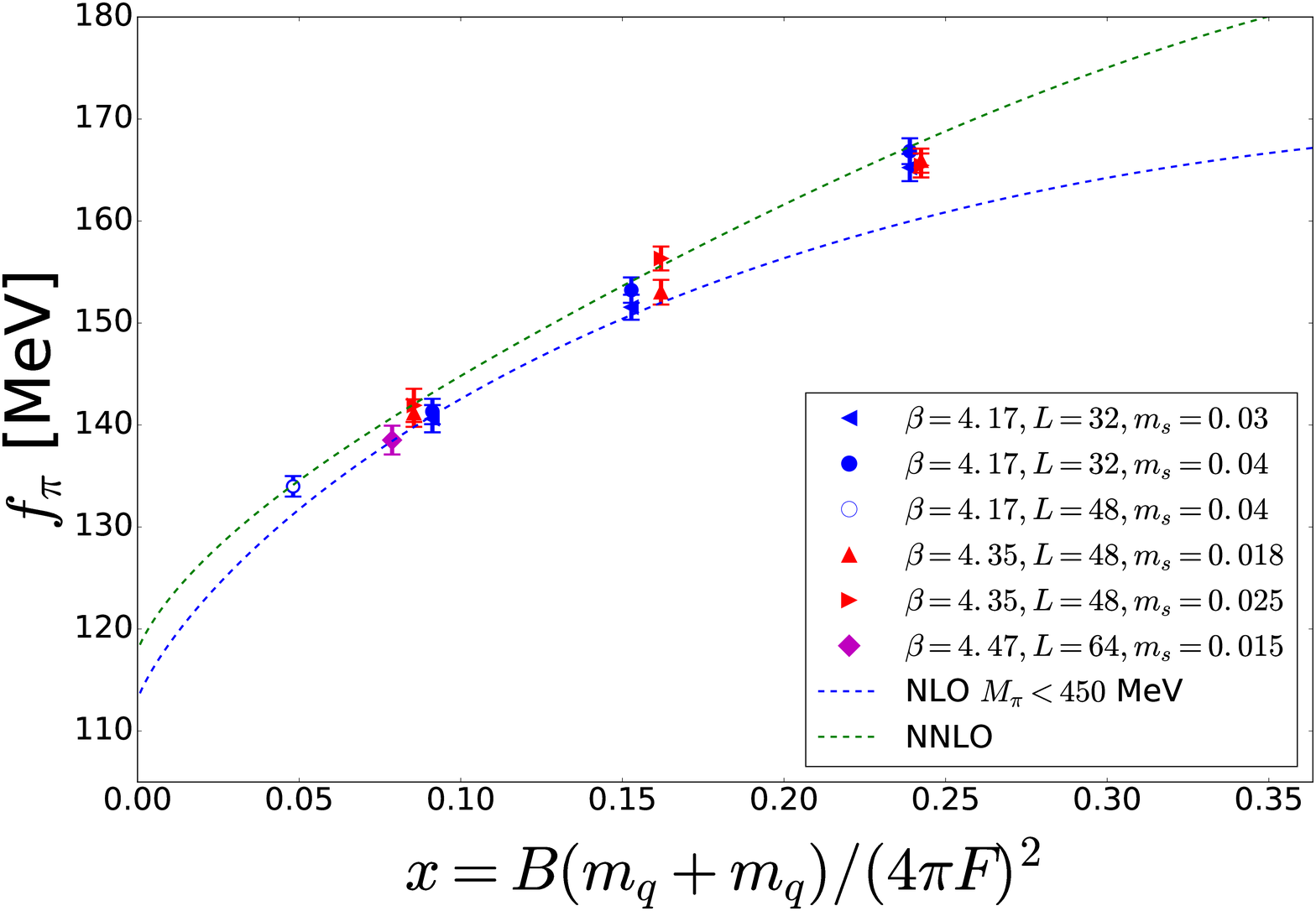}
  \caption{Plots of $M_\pi^2/\bar{m}_q $ (left panel) and $F_\pi$
    (right panel), both vs. $x=2\bar{m}_q B/(4\pi F)^2$. Fit lines
    show the best NLO (blue) and NNLO (green) fits in the continuum
    and physical strange quark mass limits. The NLO fits only include
    the ensembles for $M_\pi < 450 \text{ MeV}$}
  \label{fig:Fpi_x}
\end{figure}

To account for the strange-quark mass dependence the fit function was
corrected by a term proportional to
$M_{s\bar{s}}^2 = 2M_K^2 - M_\pi^2$. Combining with a lattice spacing
dependence, all fits were performed with a prefactor
$(1+ \gamma_{a} a^2 + \gamma_{s}(M_{s\bar{s}} -
M_{s\bar{s}}^{\text{phys}}) )$. At NLO the fits have $\chi^2$ less
than $1.5$ if including only the ensembles with pion masses
$M_\pi < 450 \text{ MeV}$, so the other ensembles were excluded for
the NLO fits. For NNLO fits the ensembles of all pion masses were
included. The results of the NLO and NNLO fits in the continuum and
physical strange quark mass limits are shown in Figure \ref{fig:Fpi_x}
by dashed lines.

\begin{figure}[t]
  \centering
  \includegraphics[width=0.49\textwidth]{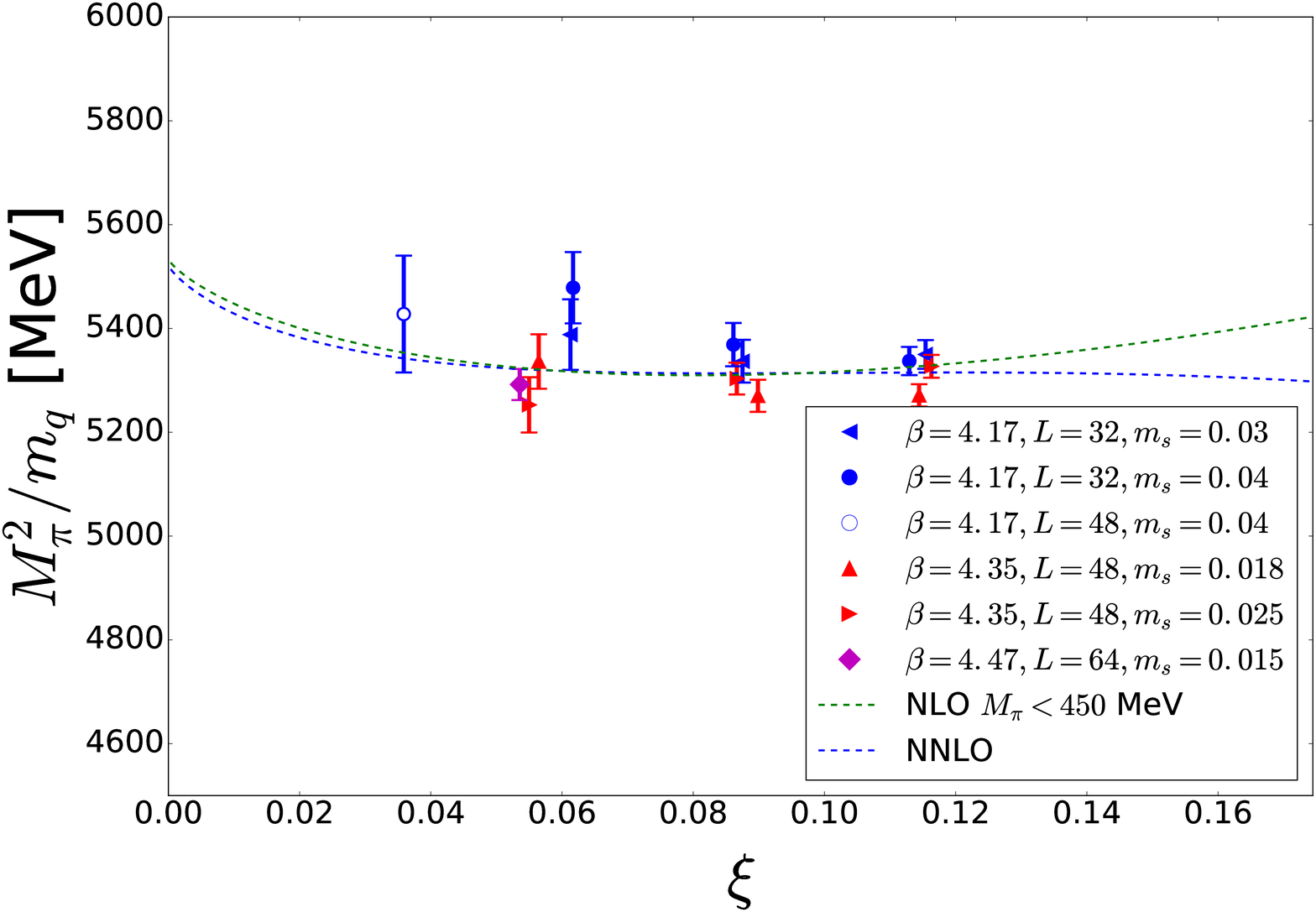}
  \includegraphics[width=0.49\textwidth]{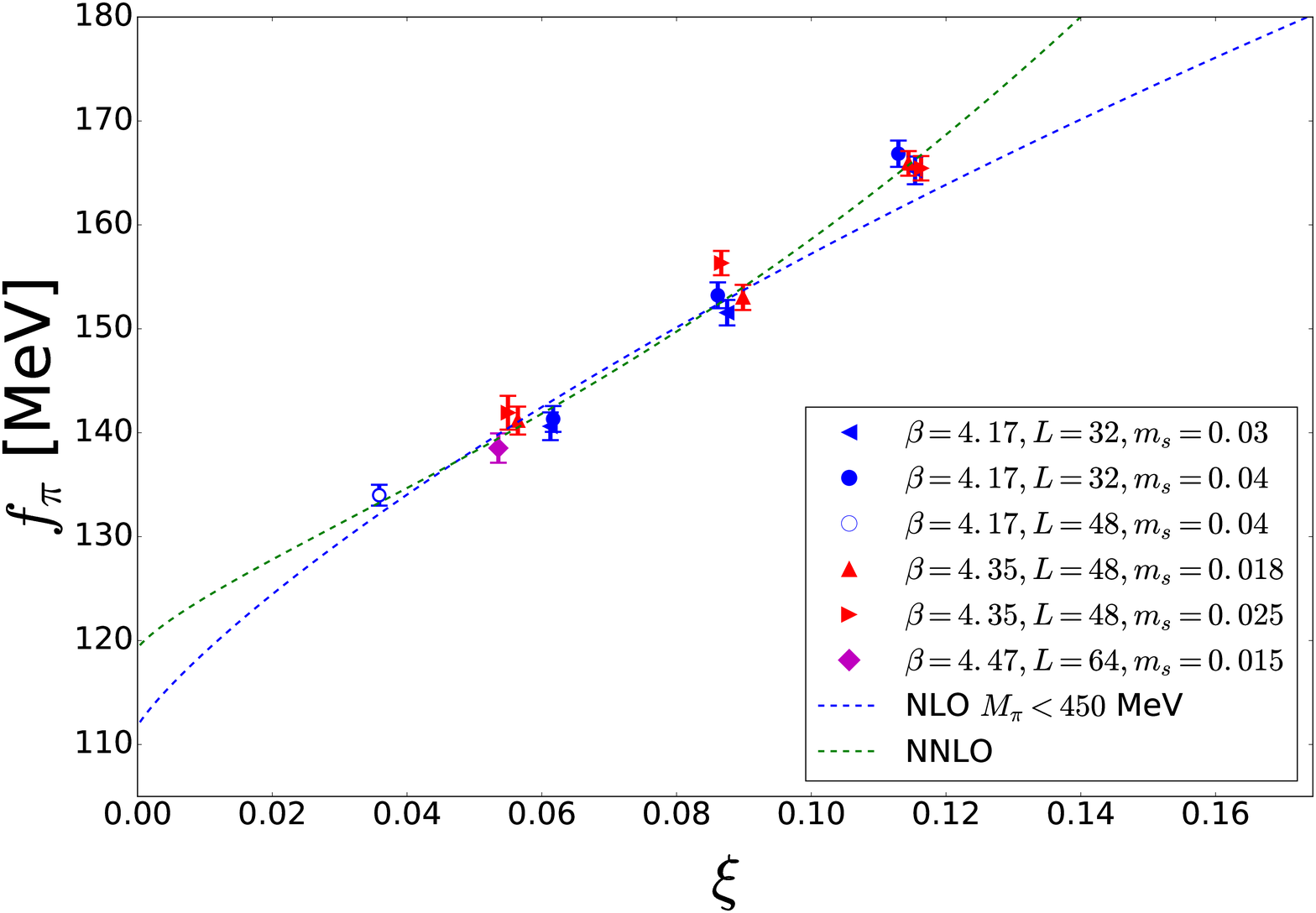}
  \caption{Same as Figure \protect\ref{fig:Fpi_x} except plotted vs. $\xi = M_\pi^2 / (4\pi F_\pi)^2$.}
  \label{fig:Fpi_xi}
\end{figure}

Alternatively the ChPT expansions can be reorganized using the
parameter $\xi = M_\pi^2 /(4\pi F_\pi)^2$. The expansions are
\begin{align}
  \frac{M_\pi^2}{\bar{m}_q } &= 2B / \s{1+\frac{1}{2} \xi \ln\frac{\Lambda_3}{M_\pi^2} - \frac{5}{8} \xi^2\p{\ln\frac{\Omega_M^2}{M_\pi^2}}^2 + c_M \xi^2 +\mathcal{O}\p{\xi^3} },\\
  F_\pi &= F / \s{1 -  \xi \ln\frac{\Lambda_4^2}{M_\pi^2}-\frac{1}{4} \xi^2\p{\ln\frac{\Omega_F^2}{M_\pi^2} }^2 + c_F \xi^2 +\mathcal{O}\p{\xi^3}},
          \label{xi-expand}
\end{align}
where similarly the values $\Omega_M$ and $\Omega_F$ are combinations
of other LEC's~\cite{PhysRevD.90.114504}. The pion masses and decay
constants are plotted against $\xi$ in Figure \ref{fig:Fpi_xi}. The
curves represent the fits of NLO and NNLO.


Our preliminary results for the LEC's from the NLO fits expanded in
$x$ are $F = 83.2(6.3)\text{ MeV}$,
$\Sigma^{1/3}[2\text{ MeV}] = 287.9(3.7)\text{ MeV}$,
$\bar{\ell}_3 = 3.11(44)$ and $\bar{\ell}_3 = 4.37(22)$. The chiral
condensate, $\Sigma$, is renormalized to the one in the
$\overline{\text{MS}}$ scheme at $\mu = 2 \text{ GeV}$ using the
renormalization factor calculated in \cite{renormalization}. The values
obtained with the two expansion parameters as well as those from NLO
and NNLO fits are all consistent within statistical error though the NNLO
results have slightly larger uncertainty. $F$ is the decay constant
in the chiral limit but at the physical pion mass value we obtain
$F_\pi = 88.9(5.2)\text{ MeV}$.



\section{Charmed mesons}

The lattice spacings of the JLQCD ensembles were chosen to treat heavy
physics with minimal cutoff effects. We produced charmed correlators
using domain-wall heavy quarks at three masses close to the charm
mass. All results shown are first interpolated to the charm mass using
the spin averaged $c\bar{c}$ masses. Charmonium correlators are also
used in the analysis of their time-moments to determine the charm
quark mass $m_c$ and strong coupling constant $\alpha_s$~\cite{charm}.

\begin{figure}[t]
  \centering
  \includegraphics[width=0.49\textwidth]{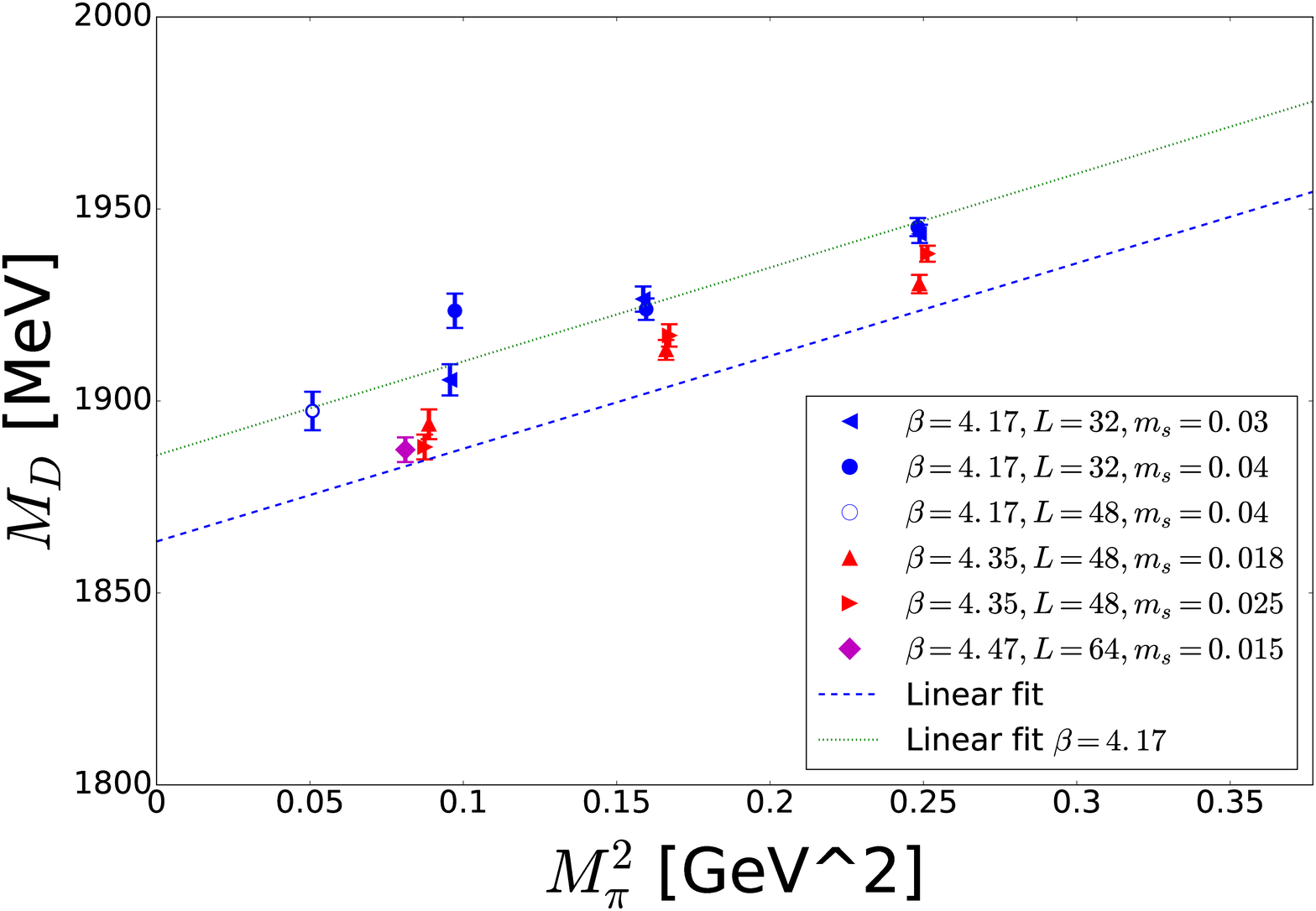}
  \includegraphics[width=0.49\textwidth]{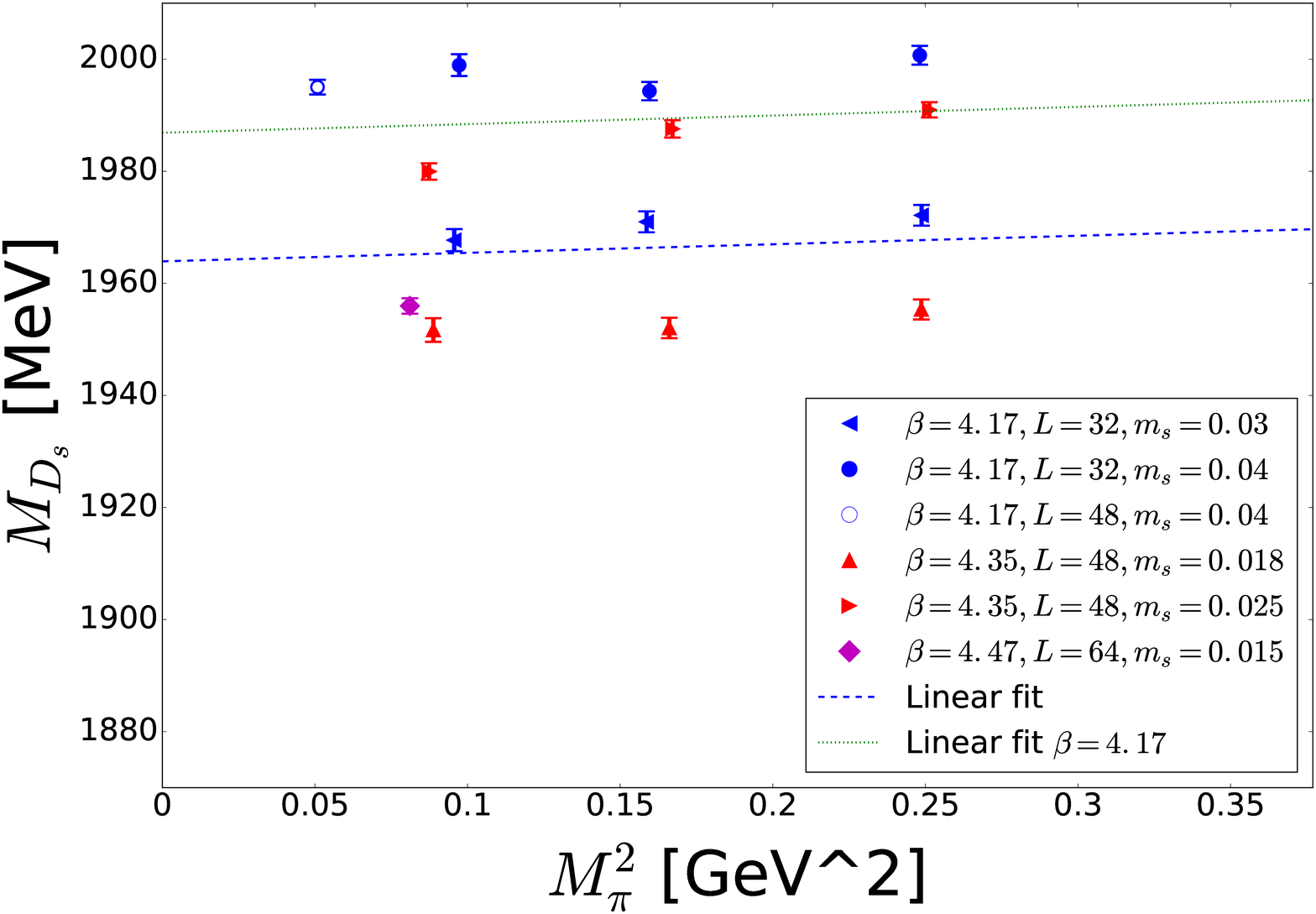}
  \caption{Masses of the $D$ meson (left) and $D_s$ meson (right) vs.
    $M_\pi^2$. These were fit linearly in $M_\pi^2$ accounting for a
    dependence on the lattice spacing $a^2$ and interpolated in the
    strange quark mass using $2M_K^2 - M_\pi^2$. The blue dashed line
    indicated the linear fit extrapolated to the continuum limit while
    the green dashed line shows the linear fit for the value of $a$
    corresponding to our coarsest lattice $\beta=4.17$.}
  \label{fig:charmedmasses}
\end{figure}

Figure \ref{fig:charmedmasses} shows the masses of the $D$ and $D_s$
mesons as well as linear fits in $M_\pi^2$ accounting for a dependence
on the lattice spacing $a^2$ and interpolated in $m_s$ using
$2M_K^2 - M_\pi^2$. The raw data for $M_{D_s}$ appear scattered
because the data points with different input strange quark masses are
plotted together. After interpolating in $2M_K^2 - M_\pi^2$, the data
at different $\beta$ are more consistent with each other. The results
after extrapolation are $M_D =1867.7(9.5)\text{ MeV}$ and
$M_{D_s} = 1964.2(5.0)\text{ MeV}$. Their experimental values are
$M_D^{\text{exp}} = 1864.8 \text{ MeV}$ and
$M_{D_s}^{\text{exp}} = 1968.3 \text{ MeV}$.  The dependence upon the
lattice cutoff $a$ turned out to be minimal with a difference of
$\mathcal{O}(1\%)$ between the fitted value at $\beta=4.17$ and the
continuum limit.

\begin{figure}[t]
  \centering
  \includegraphics[width=0.49\textwidth]{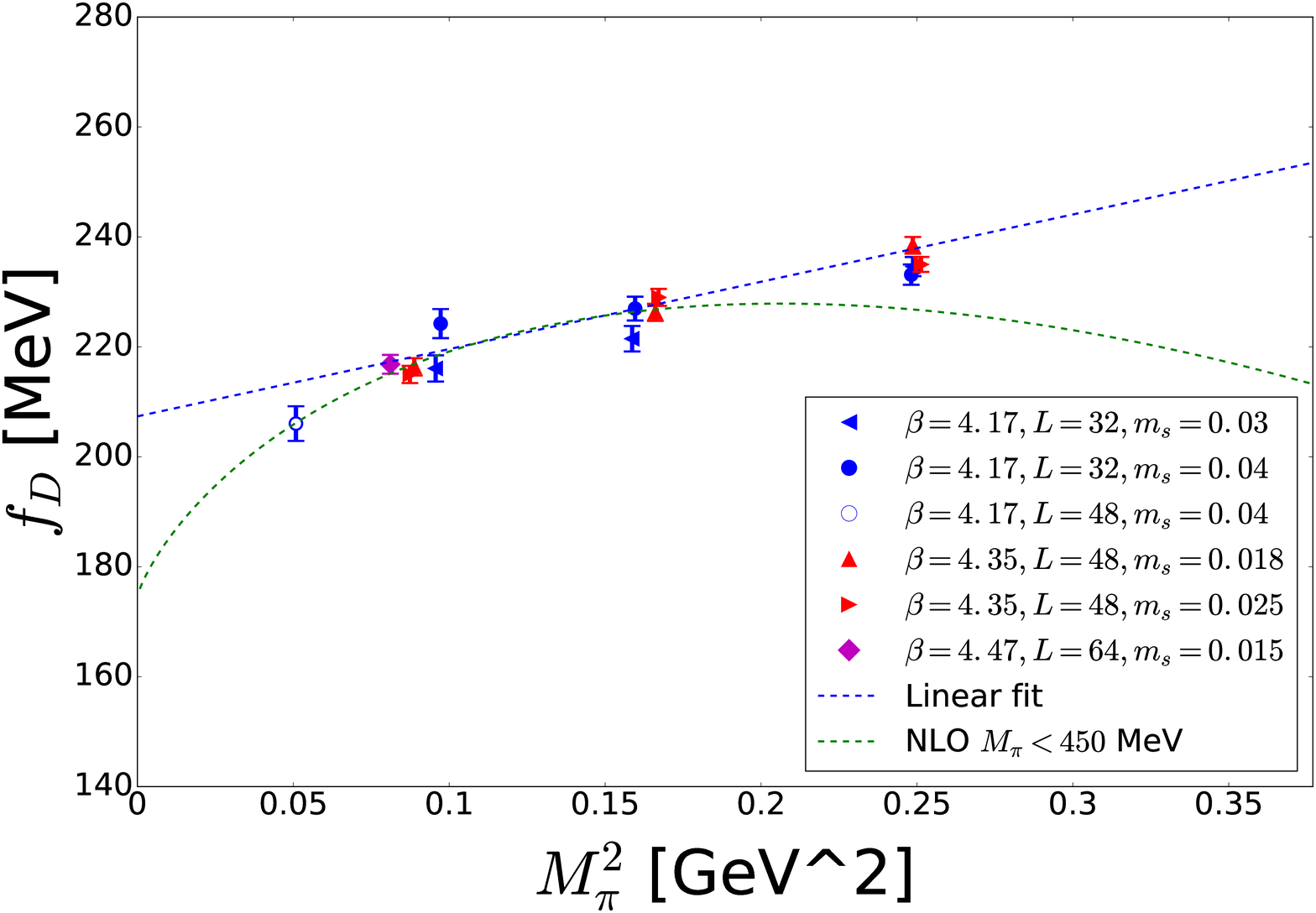}
  \includegraphics[width=0.49\textwidth]{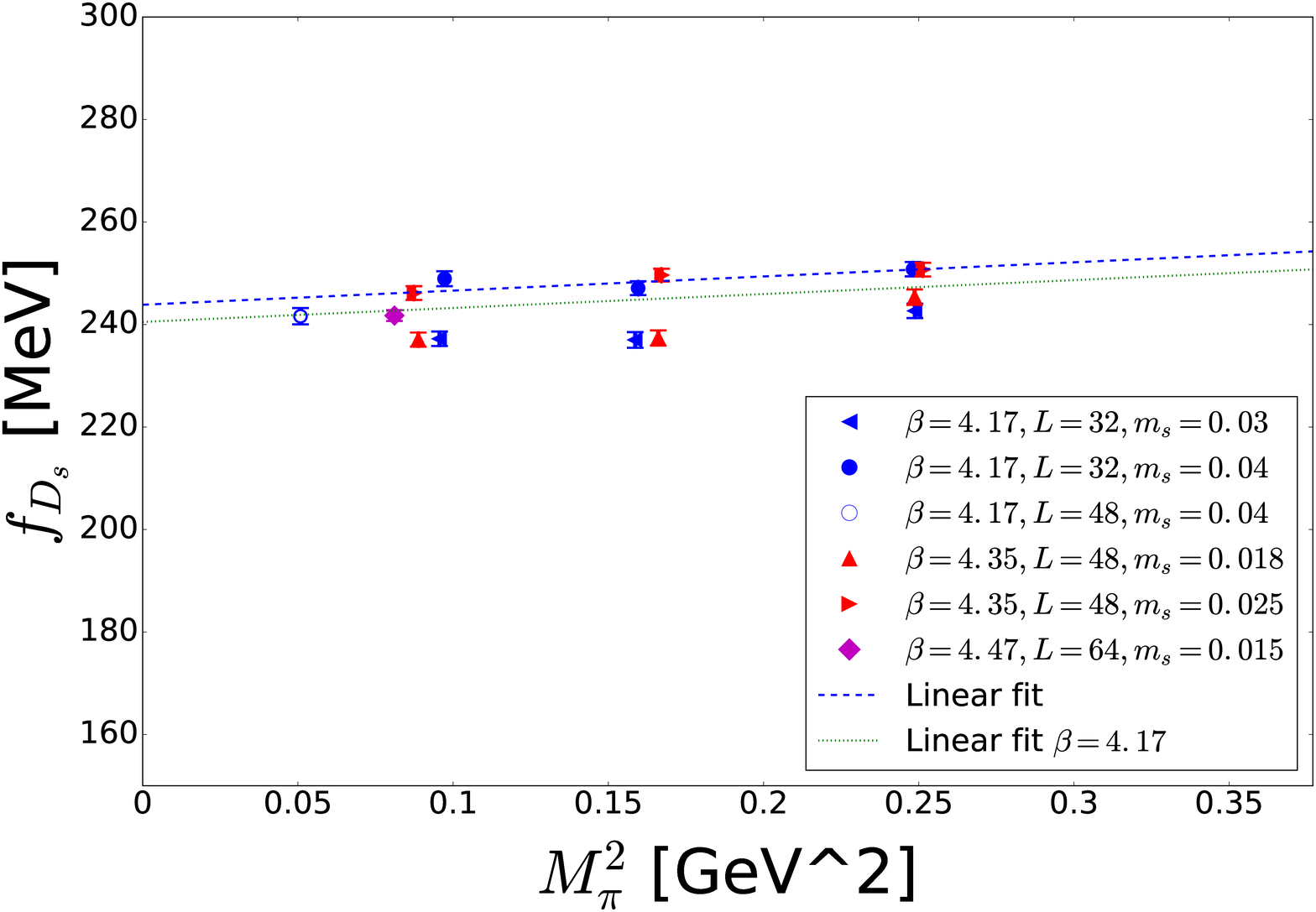}
  \caption{Charmed meson decay constants $f_D$ (left panel) and $f_{D_s}$ (right panel) vs.
    $M_\pi^2$. On both plots the blue dashed line indicate a linear
    fit in $M_\pi^2$ accounting for a dependence on the lattice
    spacing $a^2$ and interpolated in strange quark mass using $2M_K^2 -
    M_\pi^2$. The left plot of $f_D$ includes a simple linear fit as
    well as the chiral NLO fit for the ensembles with
    $M_\pi < 450 \text{MeV}$.}
  \label{fig:charmeddecay}
\end{figure}

The decay constants of the charmed mesons are also computed using the
same process for the pion using the pseudo-scalar current and the
appropriate quark masses. These results can be seen in Figure
\ref{fig:charmeddecay}. The fitted values after linear extrapolation
in $M_\pi^2$ and $a^2$ are $f_D = 209.6(5.2)\text{ MeV}$ and
$f_{D_s} = 244.4(4.1)\text{ MeV}$ with the dependence on the lattice
spacing turning out to be negligible. For the decay constant of the
$D$ meson we attempted to analyze with the ChPT fit at
NLO~\cite{Grinstein1992369} as well as a linear fit. It favors a
smaller value for $f_D$ because of the chiral logarithm, but more
precise data would be necessary to confirm, especially because the
current fit is strongly influenced by the lightest data point, which
has a relatively large error.


\section{Summary}
\label{sec:sum}

We have shown first results from the recently generated JLQCD
lattices. The good chiral properties of these lattices enable
successful fits of quantities to NLO and NNLO ChPT. Measurements are
still in progress and the precision of the decay constants and LECs
should be improved with better statistics and the use of stochastic
noise sources.

The fine lattice spacings allow us to compute charmed masses and decay
constants with small dependence upon $a$. We plan to produce results
with quarks heavier than the charm mass to investigate the lattice
spacing dependence for heavy domain wall quarks. If the dependence
continues to remain small it may be possible to extrapolate to $B$
physics.

\vspace*{0.5cm} Numerical simulations are performed on the IBM System
Blue Gene Solution at High Energy Accelerator Research Organization
(KEK) under a support of its Large Scale Simulation Program
(No. 13/14-04, 14/15-10). We thank P. Boyle for helping in the
optimization of the code for BGQ. This work is supported in part by
the Grant-in-Aid of the Japanese Ministry of Education (No. 26400259,
26247043, and 15K05065) and the SPIRE (Strategic Program for
Innovative Research) Field5 project.

\bibliography{lattice_2015_fahy}






\end{document}